\documentstyle[twoside,fleqn,npb,epsfig,amssymb]{article}
\newcommand{\AmS}{{\protect\the\textfont2
  A\kern-.1667em\lower.5ex\hbox{M}\kern-.125emS}}
\newcommand{\ycut}{\mbox{$y_{\rm cut}$}}
\newcommand{\GeV}{\, \rm GeV}
\newcommand{\etbreit}{\mbox{$E_{T, {\rm Breit}}$}}
\newcommand{\etabreit}{\mbox{$\eta_{\mbox{\scriptsize Breit}}$}}

\title{Internal Jet Structure in Dijet Production in Deep-Inelastic Scattering}

\author{Peer-Oliver Meyer\address{III. Physikalisches Institut, RWTH Aachen,
        D-52056 Aachen, Germany} for the H1 Collaboration
         }

\begin{document}

\pagestyle{empty}

\begin{abstract}
The internal jet structure in dijet production in deep-inelastic scattering
is measured with the H1 detector at HERA.
Jets with transverse energies $\etbreit > 5\GeV$ are
selected in the Breit frame employing $k_\perp$ and cone jet algorithms
in the kinematic region of squared momentum transfers of 
$10 < Q^2 \lesssim 120\GeV^2$ ~and $x$-Bjorken values of 
$2 \cdot 10^{-4} \lesssim x_{\rm Bj} \lesssim 8 \cdot 10^{-3}$.
Jet shapes and subjet multiplicities are measured as functions of 
a resolution parameter.
The corrected data are well described by QCD models.
It is observed that jets are more collimated 
with increasing transverse jet energies and decreasing pseudo-rapidities,
i.e.\  towards the photon direction.
Comparisons with OPAL data show that jet shapes of jets measured in 
$\gamma\gamma$ collisions are very similar to those 
measured in $e p$ collisions.
\end{abstract}

\hfill \parbox[t]{4cm}{\large
PITHA 99/20 \\
hep-ph/9906018
}

\vskip10mm

\begin{center}
{\bf \huge
Internal Jet Structure in Dijet Production \\ \vskip2mm
in Deep-Inelastic Scattering}

\vskip20mm

{\LARGE P.-O. Meyer \\ \vskip4mm
 \Large III. Physikalisches Institut, RWTH Aachen,\\ \vskip2mm
        D-52056 Aachen, Germany}

\end{center}

\vskip20mm

\begin{center}
\large \bf Abstract \\ \vskip1mm
\parbox[t]{14.5cm}{\large
The internal jet structure in dijet production in deep-inelastic scattering
is measured with the H1 detector at HERA.
Jets with transverse energies $\etbreit > 5\GeV$ are
selected in the Breit frame employing $k_\perp$ and cone jet algorithms
in the kinematic region of squared momentum transfers of 
$10 < Q^2 \lesssim 120\GeV^2$ ~and $x$-Bjorken values of 
$2 \cdot 10^{-4} \lesssim x_{\rm Bj} \lesssim 8 \cdot 10^{-3}$.
Jet shapes and subjet multiplicities are measured as functions of 
a resolution parameter.
The corrected data are well described by QCD models.
It is observed that jets are more collimated 
with increasing transverse jet energies and decreasing pseudo-rapidities,
i.e.\  towards the photon direction.
Comparisons with OPAL data show that jet shapes of jets measured in 
$\gamma\gamma$ collisions are very similar to those 
measured in $e p$ collisions.}
\end{center}

\vskip31mm
\begin{center}
\parbox[t]{14.5cm}{\large
Talk given on behalf of the H1 collaboration at the 7th
International Workshop on Deep-Inelastic Scattering
and QCD (DIS99), Zeuthen, April 1999.}
\end{center}

\newpage

\pagestyle{plain}
\pagenumbering{arabic}

\maketitle

\section{INTRODUCTION}

The internal structure of jets is sensitive to the mechanism by
which a complex aggregate of observable hadrons evolves from a hard
process. 
It is expected that the internal structure of jets depends mainly
on the type of the primary parton, quark or gluon, from which it originated
and to a lesser extent on the particular hard scattering process.
Measurements of the internal structure of jets have been made in
$p\bar{p}$ collisions 
and in $e^{+}e^{-}$ annihilations.
At the $e^\pm p$ collider HERA 
jet shapes have been investigated in 
photoproduction ($Q^2 \approx 0 \GeV^2$) \cite{zeus_1}
and in deep-inelastic scattering  at $Q^2 > 100\GeV^2$ \cite{zeus_2}. 

Here we present the measurements of internal jet structure \cite{h1_2}
in a sample of inclusive dijet events with transverse jet energies of
$\etbreit > 5\GeV$ in the 
kinematic range of $10 < Q^2 \lesssim 120\,\GeV^2$ and
$2 \cdot 10^{-4} \lesssim x_{\rm Bj} \lesssim 8 \cdot 10^{-3}$.
Jets are defined in the Breit frame by the $k_\perp$ \cite{seymor}
and the cone \cite{cone} jet algorithm.
The analysis is based on data taken in 1994 with the H1 detector
at HERA, operated with positrons of energy $E_{e}\,=\,27.5\GeV$
colliding with protons of energy $E_{p}\,=\,820\GeV$.
The data correspond to an integrated luminosity of 
${\cal L}_{\rm int} \simeq 2\,\mbox{pb}^{-1}$.
Two observables, jet shapes and subjet multiplicities, are studied.
Both observables are corrected for detector effects and are presented as 
a function of the transverse jet energy and the jet pseudo-rapidity.
 
\section{OBSERVABLES}

\begin{figure*}[t]
\epsfig{file=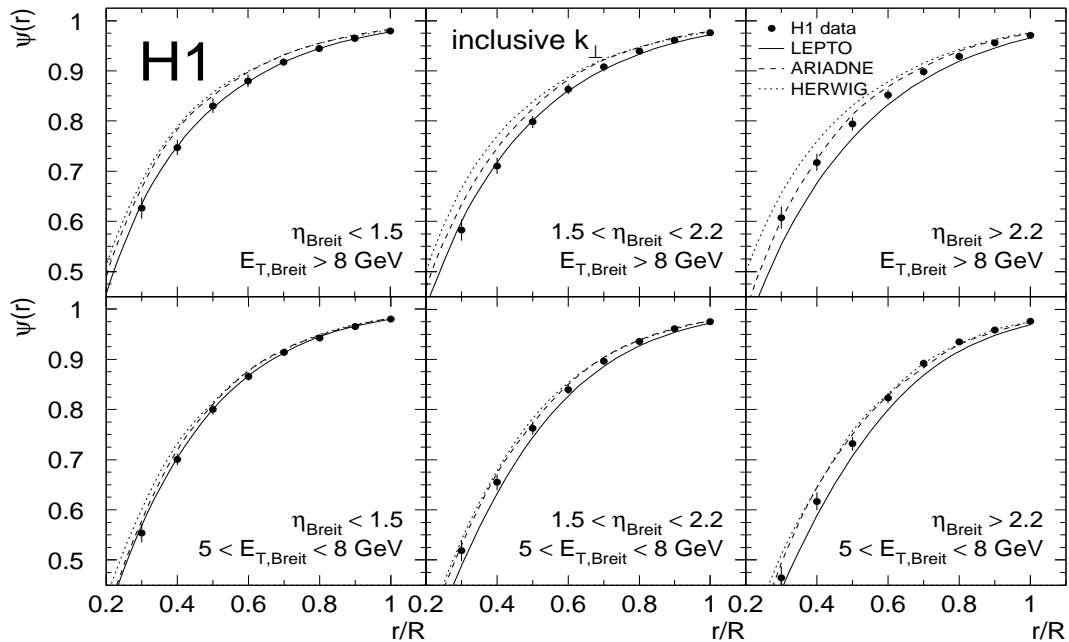,width=14.5cm,height=8.8cm}
\vskip-11mm
\caption{
The jet shapes for the inclusive $k_\perp$ algorithm.
The data are shown as a function of the transverse jet energy and the jet
pseudo-rapidity in the Breit frame (positive pseudo-rapidities
are towards the proton direction).
The results are compared to predictions of QCD models.
}
\label{fig:all}
\end{figure*}

The jet shape $\Psi(r)$ is defined as the average fractional transverse jet
energy that lies in a subcone of radius $r$ concentric with the jet axis
\begin{equation}
\Psi(r) \equiv \frac{1}{N_{\rm jets}} \sum_{\rm jets} \;
\frac{E_{T}(r)}{E_{T, {\rm jet}}} \, .
\label{eq:defpsi}
\end{equation}
$N_{\rm jets}$ is the total number of jets in the sample and 
$E_{T}(r)$ is the transverse energy within a subcone of radius $r$.
A natural variable for studying the internal structure of jets with the
$k_\perp$ cluster algorithm is the multiplicity of subjets, resolved at 
a resolution scale which is a fraction of the jet's transverse energy.
For each jet in the sample
the clustering procedure is repeated for all particles assigned
to the jet.
The clustering is stopped when the distances $y_{ij}$ between
all particles $i,j$ are above some cut-off \ycut
\begin{equation}
y_{ij} = \frac{\min (E_{T,i}^2 , E_{T,j}^2)}{E_{T, {\rm  jet}}^2} 
\frac{\left( \Delta \eta_{ij}^2 + \Delta \phi_{ij}^2 \right)}{R_0^2}
>\ycut 
\end{equation}
where $R_0$ is set to $R_0=1.0$.
The remaining (pseudo-)particles are called subjets.
The parameter \ycut ~defines the minimal relative transverse
energy between subjets inside the jet and thus determines the extent 
to which the internal jet structure is resolved.

\section{RESULTS}

The radial dependence of the jet shape $\Psi(r)$ for 
the $k_\perp$ algorithm is shown in Fig.~\ref{fig:all} 
in different ranges of the pseudo-rapidity and transverse jet energy
in the Breit frame.
The predictions of three QCD models are overlaid.
The jet shape $\Psi(r)$ increases faster with $r$ for jets
at larger $E_T$, indicating that these jets are more collimated.
Jets towards the proton direction (at larger values of \etabreit)
are broader than jets towards the photon direction 
(smaller \etabreit).

The QCD models LEPTO, ARIADNE and HERWIG all show \etbreit ~and 
\etabreit ~dependencies similar to those seen in the data.
LEPTO gives a good overall description of the data although 
it has the tendency to produce broader jets in the proton direction.
A good description is also obtained by the ARIADNE model
except for jets at smaller pseudo-rapidities where the jet shapes
have the tendency of being too narrow.
For the HERWIG model the jet shapes are slightly narrower than those
in the data in all \etbreit ~and \etabreit ~regions.

\begin{figure}[ht]
  \epsfig{file=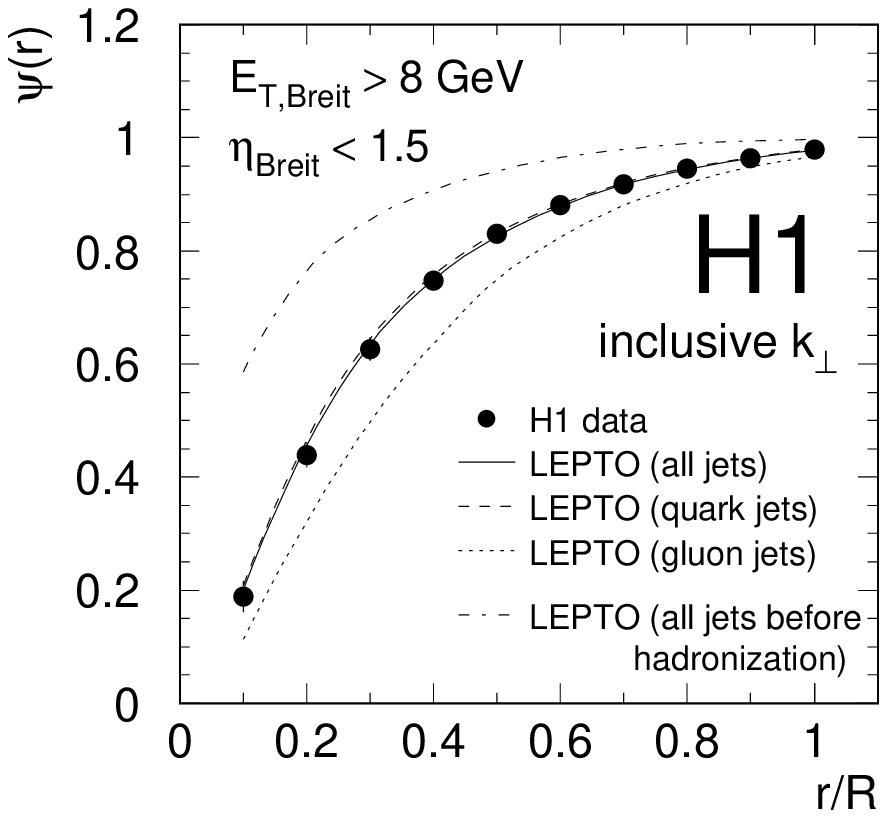,width=7.0cm,height=4.8cm}
  \epsfig{file=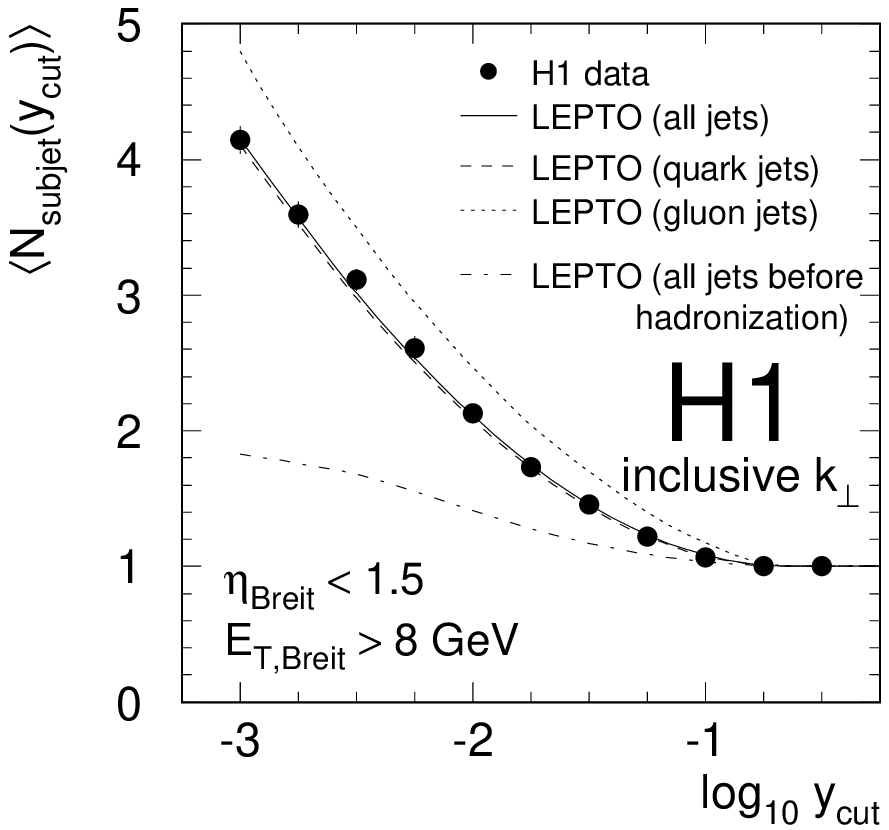,width=7.0cm,height=4.8cm}
\vskip-9mm
\caption{
Model predictions of the internal structure of quark and gluon jets
for the inclusive $k_\perp$ algorithm by the LEPTO parton shower model.
The jet shapes (top) and the subjet multiplicities (bottom)
are shown separately for quark and gluon induced jets with 
$\etbreit > 8\,{\rm GeV}$ and $\eta_{\rm Breit} < 1.5$,
together with the sum of both and the comparison to the H1 measurement.
The distributions of the observables before hadronization are also shown.
}
\label{fig:models}
\end{figure}

Fig.~\ref{fig:models} shows the jet shapes and the 
subjet multiplicities for the $k_\perp$ algorithm 
as predicted by the LEPTO
parton shower model, separately for
quark and gluon jets at $\etbreit > 8\,{\rm GeV}$ and $\etabreit < 1.5$.
Within this model gluon jets are broader than quark jets.
The same prediction is obtained by the HERWIG parton shower model.
In the phase space considered here, LEPTO and HERWIG (in agreement with 
next-to-leading order calculations) predict a fraction of approx.\ 
80\,\% photon-gluon fusion events with two quarks in the partonic final
state.
The jet samples of these models are therefore dominated by quark jets.
Both model predictions for the jet shapes and the subjet
multiplicities therefore mainly reflect the properties of the 
quark jets as can be seen in Fig.~\ref{fig:models}.
These predictions give a good description of the data.
Thus, we conclude, that the jets we observe are consistent with 
being mainly initiated by quarks.

\begin{figure}[t]
\epsfig{file=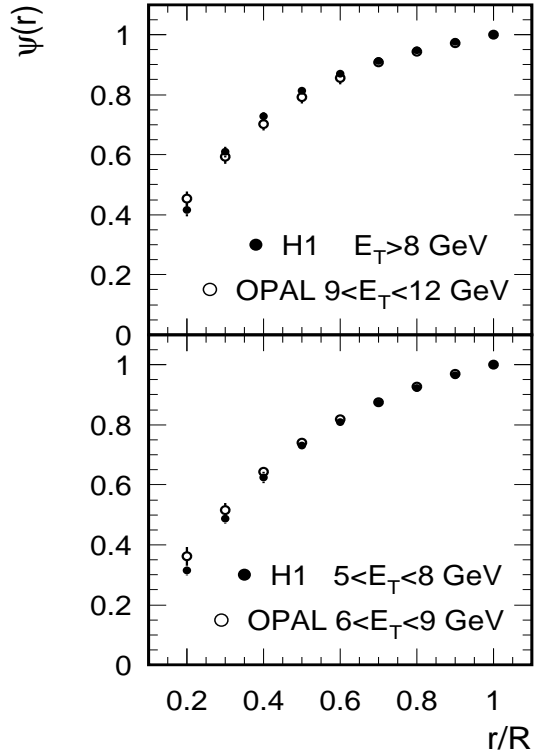,width=7.5cm,height=10.5cm}
\vskip-11mm
\caption{
Jet shapes for the cone algorithm at $\etabreit < 1.5$. 
The jet shapes of jets measured by OPAL in $\gamma\gamma$ collisions 
are overlaid.
The comparison is shown for two different regions of
transverse jet energies. 
}
\label{fig:opal}
\end{figure}

Fig.~\ref{fig:opal} shows the jet shapes for the cone algorithm 
in the backward region ($\etabreit < 1.5$) for two regions 
of $\etbreit$.
As seen for the $k_\perp$ algorithm in Fig.~\ref{fig:all} jets with
larger transverse energy $\etbreit$ are more collimated.

The results are compared to the jet shapes 
measured by OPAL in $\gamma\gamma$ collisions
\cite{opal} in similar $E_{T}$ regions.
Although these jets are produced in a different scattering process,
their jet shapes are very similar to those measured in $e p$ collisions.
This indicates that the internal jet structure does not depend on
the underlying hard process.


\begin{thebibliography}{9}
\bibitem{zeus_1} ZEUS Collaboration, J. Breitweg et al.,
  Eur. Phys. J. {\bf C2}
(1998) 61.
\bibitem{zeus_2} ZEUS Collaboration, J. Breitweg et al.,
DESY 98-038, Hamburg, Germany (1998)
\bibitem{h1_2} H1 Collaboration, C. Adloff et al., Nucl. Phys. {\bf B545} (1999) 3.
\bibitem{seymor} S.D. Ellis, D.E. Soper, Phys. Rev. {\bf D48} (1993) 3160.
\bibitem{cone} L.A. del Pozo, PhD Thesis University of Cam\-bridge, 
Cambridge, England (1993) RALT-002
\bibitem{opal} OPAL Collaboration, G. Abbiendi et al., CERN-EP/98-113.
\end{thebibliography}
\end{document}